%% file: main.tex
\documentclass[a4paper,11pt]{article}

\usepackage{jinstpub} 
\usepackage[table]{xcolor}
\usepackage{booktabs}
\usepackage{multicol}
\usepackage{multirow}
\usepackage{lineno}

\usepackage{listings}
\definecolor{codegreen}{rgb}{0,0.6,0}
\definecolor{gray}{rgb}{0.5,0.5,0.5}
\definecolor{codepurple}{rgb}{0.58,0,0.82}

\lstdefinestyle{myCPPStyle}{
  language=C++,
  aboveskip=3mm,
  belowskip=3mm,
  showstringspaces=false,
  columns=flexible,
  basicstyle={\small\ttfamily},
  numbers=none,
  numberstyle=\scriptsize\color{gray},
  keywordstyle=\color{blue},
  commentstyle=\color{codegreen},
  stringstyle=\color{magenta},
  breaklines=true,
  breakatwhitespace=true,
  tabsize=2
}

\usepackage{xspace}
\input{commands}

\proceeding{3$^{\text{rd}}$ Artificial Intelligence for the Electron Ion Collider workshop -- AI4EIC2023\\
November 28 - December 1, 2023\\
Catholic University of America, Washington (D.C.), USA}

\title{\boldmath Particle identification with machine learning from incomplete data in the ALICE experiment}

\collaboration[c]{on behalf of the ALICE collaboration}

\author[a,b,1]{Maja Karwowska,\note{Corresponding author.}}
\author[a]{{\L}ukasz Graczykowski}
\author[c, d]{Kamil Deja}
\author[c]{Miłosz Kasak}
\author[a]{and Małgorzata Janik}
\affiliation[a]{Faculty of Physics, Warsaw University of Technology\\Koszykowa 75, 00-662 Warsaw, Poland}
\affiliation[b]{CERN -- European Organization for Nuclear Research\\Espl. des Particules 1, 1211 Geneva, Switzerland}
\affiliation[c]{Faculty of Electronics and Information Technology, Warsaw University of Technology\\Nowowiejska 15/19, 00-665 Warsaw, Poland}
\affiliation[d]{IDEAS NCBR\\Chmielna 69, 00-801 Warsaw, Poland}

\emailAdd{maja.karwowska@cern.ch}

\abstract{%
The ALICE experiment at the LHC measures properties of the strongly interacting matter formed in ultrarelativistic heavy-ion collisions. Such studies require accurate particle identification (PID). ALICE provides PID information via several detectors for particles with momentum from about 100 \MeVc~up to 20 \GeVc.
Traditionally, particles are selected with rectangular cuts. A~much better performance can be achieved with machine learning (ML) methods. Our solution uses multiple neural networks (NN) serving as binary classifiers. Moreover, we extended our particle classifier with Feature Set Embedding and attention in order to train on data with incomplete samples. We also present the integration of the ML project with the ALICE analysis software, and we discuss domain adaptation, the ML technique needed to transfer the knowledge between simulated and real experimental data.
}


\begin{document}
\maketitle
\flushbottom

\section{Introduction}
\label{sec:intro}

\begin{figure}[b]
    \centering
    \includegraphics[width = .7\textwidth]{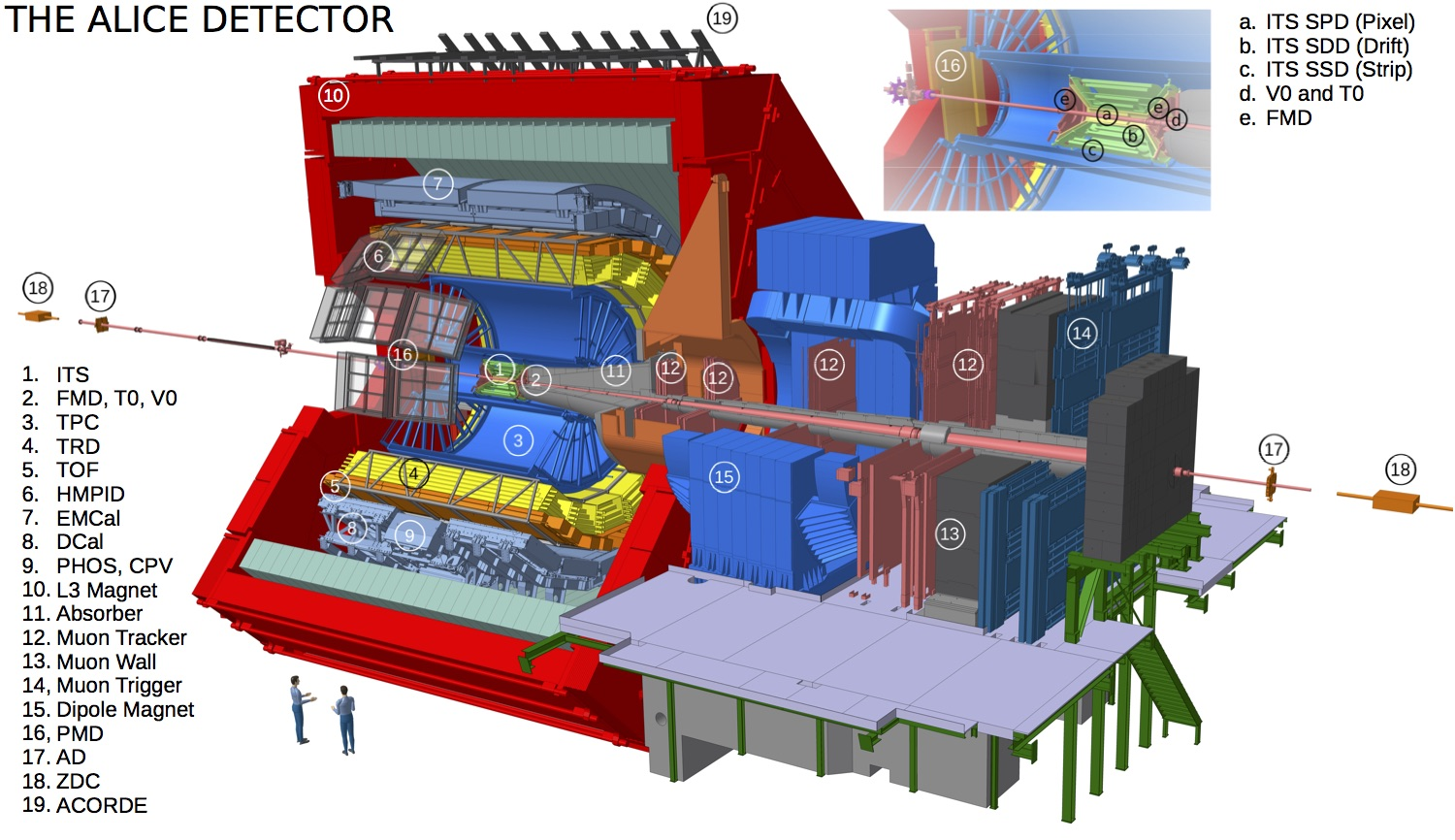}
    \caption{Components of the~ALICE detector in its Run 2 configuration \cite{Tauro:2263642}.}
    \label{fig:alice-detectors}
\end{figure}

ALICE (A Large Ion Collider Experiment)~\cite{Aamodt:2008zz} is one of the four major detectors located at the Large Hadron Collider~(LHC) at CERN~\cite{Evans:2008zzb}. The main goal of ALICE is to measure the properties of the quark--gluon plasma (QGP), a deconfined state of quarks and gluons, theorized to exist in the early Universe~\cite{ALICE:2022wpn}. Detailed studies of QGP require very precise particle identification (PID), i.e., the ability to discriminate between different particle species produced during the collision. High PID accuracy distinguishes ALICE from other LHC experiments. It also allows for selecting a subset of particles required for specific analysis. Thanks to several detectors operating concurrently, various types of particles can be separated over a wide range of momentum from just around 100~\MeVc~up to around 10~\GeVc. Figure \ref{fig:alice-detectors} presents a scheme of the ALICE detectors as used during the Run 1 and Run 2 LHC data-taking periods.

In particular, particle identification over the full azimuthal angle uses information from three detectors: Time Projection Chamber~\cite{tpc-tdr} (TPC), Time-of-Flight~\cite{tof-tdr} (TOF), Transition Radiation Detector~\cite{trd-tdr} (TRD). The TPC is one of the most important ALICE detectors as it records the 3D trajectory of charged particles. It also measures particle-specific ionization energy loss, which is essential for PID. The TOF takes measurements of particle time of flight from the collision vertex to the detector. Particle velocity and mass can be further computed from the time of flight and the track information. The TRD records transition radiation, the emission of photons by electrons traversing the boundaries of a radiator. It enables to distinguish electrons from other charged particles. Since all the aforementioned detectors detect particles carrying a non-zero electric charge, our work focuses on the identification of charged particles.

The particles are identified based on observables derived from signals recorded by the detectors using selection criteria compared with theoretical calculations such as the Bethe-Bloch formula in case of TPC signals. The traditional approach, the so-called $\rm{n}_{\sigma}$ method, is to compare the number of standard deviations from the expected value for all detector-specific observables. If a particle has associated PID observables exceeding a certain number of standard deviations, it is rejected. For instance, if one intends to use both TPC and TOF signals, the PID selection would be defined as: $\rm \sqrt{n_{\sigma,TPC}^{2}+n_{\sigma,TOF}^{2}}<\Lambda$, where $\Lambda$ depends on desired balance between purity and efficiency, and typically is in range of 2 to 3. Such an approach is justified when the separation between various particle species is significantly large. However, in reality, the characteristics of different particle species can overlap, and combining information from multiple detectors can become very complex.

\section{PID with machine learning}
\label{sec:pid-ml}

A natural response to the difficulties with optimal particle selection is to use machine learning (ML) algorithms, particularly the Bayesian method and neural networks (NN). In this view, particle identification is a standard classification problem. Compared to a human analyzer, ML can utilize more input particle features and learn more complex relationships between the variables. The Bayesian approach~\cite{ALICE:2016zzl} is available in the new ALICE software framework O$^2$, but its flexibility is limited. For this reason, we focused on neural networks,  whose usage for particle identification was explored only for very specific analysis cases in other experiments~\cite{lhcb2015lhcb, collado2021learning, cms2022identification}.

\subsection{Neural network approach}
\label{sec:nn}

The simplest model, and also the starting point of our analysis, is a single feed-forward network trained and applied on Monte Carlo simulated data. We implemented a binary classifier, one instance per each (anti)particle species and each combination of detector signals. The network outputs a single value normalized to the range $(0, 1)$ by applying the logistic function $f(x) = \frac{1}{1+e^{-x}}$. The output value corresponds to the probability of the example corresponding to a specific particle type based on a selected detector set of measurements (TPC only, TPC+TOF, TPC+TOF+TRD). It is not possible to process different combinations of detector signals by a single network because the choice of detector set impacts the size of the network input vector. The networks for each (anti)particle species and detector setup are trained independently. Initial results reported in Ref.~\cite{JINST2022} are promising: a~simple network using combined information from TPC and TOF detectors can improve efficiency compared to the traditional method while not losing purity.

Nevertheless, it must be considered that it is more difficult to estimate the systematic uncertainties of machine learning models. An example solution is to use dropout, a standard method for reducing overfitting in neural networks, to approximate Bayesian uncertainty as shown in~\cite{gal2016dropout}.

\subsection{Integration of PID ML with the \osq framework}
\label{sec:o2-pid}

The ALICE analysis framework, \osq~\cite{o2physics}, is written in C++ and heavily utilizes the ROOT~\cite{BRUN199781} library, while most popular machine learning frameworks are implemented in Python. Therefore, we needed to build a universal interface between Python-based machine learning projects and the C++-based analysis framework. Our solution makes use of the ONNX (Open Neural Network Exchange) standard~\cite{onnx}, which defines a common file format for storing machine learning models developed in various frameworks such as Tensorflow~\cite{tensorflow2015-whitepaper} and PyTorch~\cite{pytorch}. The ONNXRuntime~\cite{onnxruntime} can then be used for ONNX model inference in C++.

\begin{figure}
    \centering
    \includegraphics[width = \textwidth]{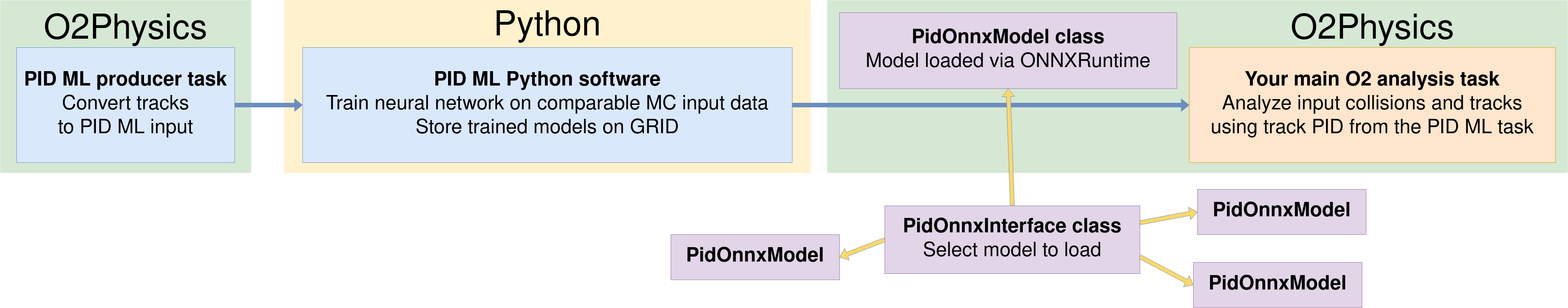}
    \caption{Scheme of the current ML particle identification workflow in O$^2$. More information about the workflow topology in the O$^2$ Analysis Framework can be found in~\cite{Alkin:2021mfo}.}
    \label{fig:pidml-interface}
\end{figure}

The current PID ML workflow in \osq is presented in Figure~\ref{fig:pidml-interface}. The processing starts from the analysis input data: reconstructed collisions and tracks in AO2D files. For training, simulated Monte Carlo data is used as it contains particle species labels. The PID ML producer task filters the input with a few rough preselections to exclude meaningless tracks and produces skimmed data that contains only track properties used by a neural network. The skimmed data is used for model training. Trained models are stored in the ONNX format in the Condition and Calibration Data Base (CCDB), which is accessible by analyses running on the worldwide LHC computing GRID. In an analysis task, a PID ML model for a single particle and detector settings is represented by an instance of the PidOnnxModel class. In more complex use cases, one can use PidOnnxModelInterface instead of dealing with a collection of PidOnnxModel instances. PidOnnxModelInterface provides handy management of multiple ML models, each with different target particle species, detector setup, and acceptance threshold.

Utilization of ONNXRuntime comes at the cost of creating additional memory copies. The input data is stored on disk in ROOT file format and read as Apache Arrow tables. Unfortunately, no direct conversion exists between Arrow tables and ONNXRuntime tensors. Therefore, the data of interest needs to be copied into C++ STL vectors and then converted to a tensor. This requires hardcoding of inputs for each ML model, which results in substantial code repetition in all analyses using ONNXRuntime and prevents the full development of a universal ONNXRuntime interface in \osq. The aforementioned problems are visible in the inference example in Listing~\ref{lst:current-code}.
\clearpage

\begin{lstlisting}[language=C++, captionpos=b,
basicstyle={\scriptsize\ttfamily},
caption={An example of model inference in \osq with ONNXRuntime.},
label={lst:current-code}]
std::vector<float> inputValues{track.px(), track.py(), track.pz(), ...);
if (mDetector >= kTPCTOF) inputValues.push_back(scaledTOFSignal);
std::vector<Ort::Value> inputTensors;
inputTensors.emplace_back(Ort::Experimental::Value::CreateTensor<float>(inputValues.data(), inputValues.size(), input_shape));
auto outputTensors = mSession->Run(mInputNames, inputTensors, mOutputNames);
\end{lstlisting}

There exists at least one alternative to ONNXRuntime, ROOT SOPHIE TMVA. It is being developed as part of the ROOT library and enables inference of neural networks saved in ONNX file format. SOPHIE is naturally adapted to the ROOT file format, making it easier to adapt to the ROOT-Arrow \osq data pipeline. Nevertheless, SOPHIE is still in the development stage. The framework lacks the implementation of many advanced neural network operators as well as the implementation of other ML models such as Boosted Decision Trees used by various groups in ALICE.

Therefore, the main effort is still organized around a more analysis-friendly adaptation of ONNXRuntime. Once a direct, preferentially copyless Arrow-ONNXRuntime conversion is developed, it will be possible to use the so-called IO Binding to map inputs between data and an ML model. It will also enable encoding model inputs as, for example, strings in a JSON configuration file. This will result in a universal, analysis-independent inference code as depicted in Listing~\ref{lst:desired-code}.

\begin{lstlisting}[language=C++, captionpos=b,
basicstyle={\scriptsize\ttfamily},
caption={Proposed code of model inference with ONNXRuntime without input hardcoding.},
label={lst:desired-code}]
std::vector<std::string> inputNamesFromJson = readJson(jsonFileName);
for (std::string& name : inputNamesFromJson)
    mBinding->BindInput(name, table.asArrowTable()->GetColumnByName(name));
Ort::MemoryInfo outputMemInfo{"Cpu", OrtDeviceAllocator, 0, OrtMemTypeDefault};
mBinding->BindOutput("output", outputMemInfo);
mSession->Run(Ort::RunOptions(), *mBinding);
auto outputTensors = mBinding->GetOutputValues();\nopagbreak[4]
\end{lstlisting}

Presently, the same integration design that PID ML has, with ONNXRuntime and a C++ class to wrap management of a single ONNX model, was applied in other machine learning projects in \osq. A common interface that further unified various implementations of the ONNX integration was developed based on PidOnnxModel. It is already used for supporting calculations for TPC PID and selection of $\Lambda_c^+$ in the $\rm{pK}^-\pi^+$ decay channel. A specialization of the unified interface is being propagated to all heavy-flavor analyses. All of these developments encountered the same difficulties with input hardcoding and code repetitions, and they would benefit from code simplification brought by Arrow-ONNXRuntime direct conversion.

\section{Feature Set Embedding and the attention mechanism}
\label{sec:fse-attention}

Even though an analyzer can set his own choice of detector set, it does not ensure that all input samples contain all needed detector signals. Particle properties are measured independently by each ALICE detector. Therefore, a particle can be recorded by a subset of detectors while not being measured by others. Possible reasons can be random, like detector malfunction or switching off, and systematic, like particle properties falling outside the detector acceptance region, e.g., too low transverse momentum, \pt, to reach outer detectors like TOF and TRD.

One can simply remove all incomplete samples from the input. However, this does not allow the identification of samples with missing information, which can form the majority of data. Another method is imputation, which introduces artificial bias to the initial dataset, which can disturb the predictions of the ML algorithm. It is also possible to alter the neural network architecture. For example, one can use a neural network ensemble, a set of classifiers, one per each subset of the training dataset without missing data. The major drawback of this approach is computational complexity, especially with a growing set of attributes with missing values.

\begin{figure}
    \centering
    \includegraphics[width=\textwidth]{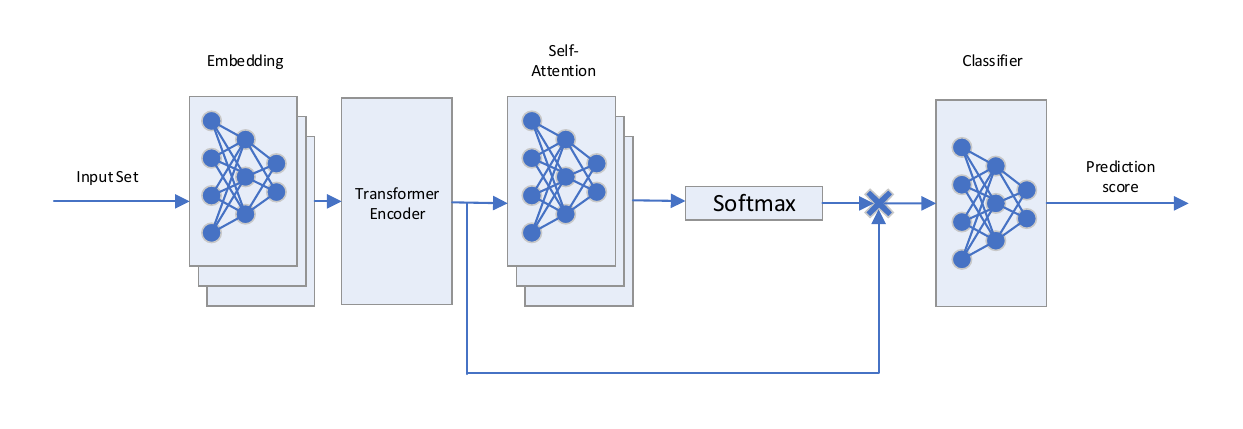}
    \caption{The proposed model architecture. Layered blocks are applied separately to each vector in a set. Single blocks are applied to their input as a whole.} \label{fig:model_arch}
\end{figure}

To overcome the aforementioned shortcomings, in Ref.~\cite{kasak2023machinelearningbased}, we introduce a novel method based on the attention mechanism, similar to the method introduced for a medical use-case in AMI-Net~\cite{wang2019attention}. The system overview is shown in Figure~\ref{fig:model_arch}.

The first module is based on the Feature Set Embedding strategy proposed in Ref.~\cite{grangier2010feature}. Input data samples are encoded into a set of feature-value pairs. Each pair represents a non-missing value in the input sample and a one-hot encoded index of the feature corresponding to this value. Then, a neural network with a single hidden layer computes embedding for each feature-value pair. The embeddings place similar features close in the embedded space.

The Transformer~\cite{vaswani2017attention} encoding module connects different features represented by a set of embedding vectors and finds input patterns. For example, a detector signal has meaning only if the momentum is within a particular range. The softmax function is applied to the attention output, a variable-size set of vectors. An additional self-attention layer is used to merge these vectors into one. Finally, the pooled vector is processed by the simple classifier described in Section~\ref{sec:nn}.

The attention architecture was tested on data from a Monte Carlo simulation of proton--proton collisions at $\sqrt{s}=13$~TeV with a realistic simulation of the time evolution of the detector conditions in the LHC Run 2 data-taking period. The simulation was performed with Pythia8~\cite{Sjostrand:2014zea}, the Geant~4~\cite{Brun:1994aa} particle transport model, and general-purpose settings. The six most abundant particle species were considered for comparison: pions, kaons, protons, and their antiparticles.

The results reported in Ref.~\cite{kasak2023machinelearningbased} clearly show that machine learning algorithms easily outperform the standard method as measured by $F_1$ metrics. The proposed attention architecture achieves very high scores of $F_1$, precision (purity), and recall (efficiency), comparable with other analyzed ML models. At the same time, our model avoids the flaws of other solutions: artificial bias in imputed and case-deleted data and potentially larger complexity of the neural network ensemble.

\section{Domain Adversarial Neural Networks}
\label{sec:domain-adaptation}

Particle identification is used to discriminate particle species in both real experimental data and Monte Carlo simulations. In particular, machine learning techniques presented in this article learn on labeled simulated data but can also be applied to unlabeled experimental data. However, the simulations often result in distributions of particle features shifted as compared to values registered at the experiment. To mitigate this effect, standard PID methods utilize automated processes for data domain alignment. For example, ALICE implemented a tuning method of simulated signals, which shifts back simulated distributions to reproduce, on average, the collected data distributions of selected variables.

Naturally, this is a limited solution, which does not allow for full domain alignment of all particle features. Therefore, we will make use of a known machine learning technique called domain adaptation, which aims to learn the discrepancies between two data domains, the labeled source and the unlabeled target, and translate those to a single hyperspace. The desired classifier is trained and applied to features from the combined latent space. Since the classifier works independently of the initial data domains, it achieves similar performance on both the labeled and unlabeled data (simulated and experimental data in our case).

\begin{figure}[hbt]
    \centering
    \includegraphics[width=0.6\textwidth]{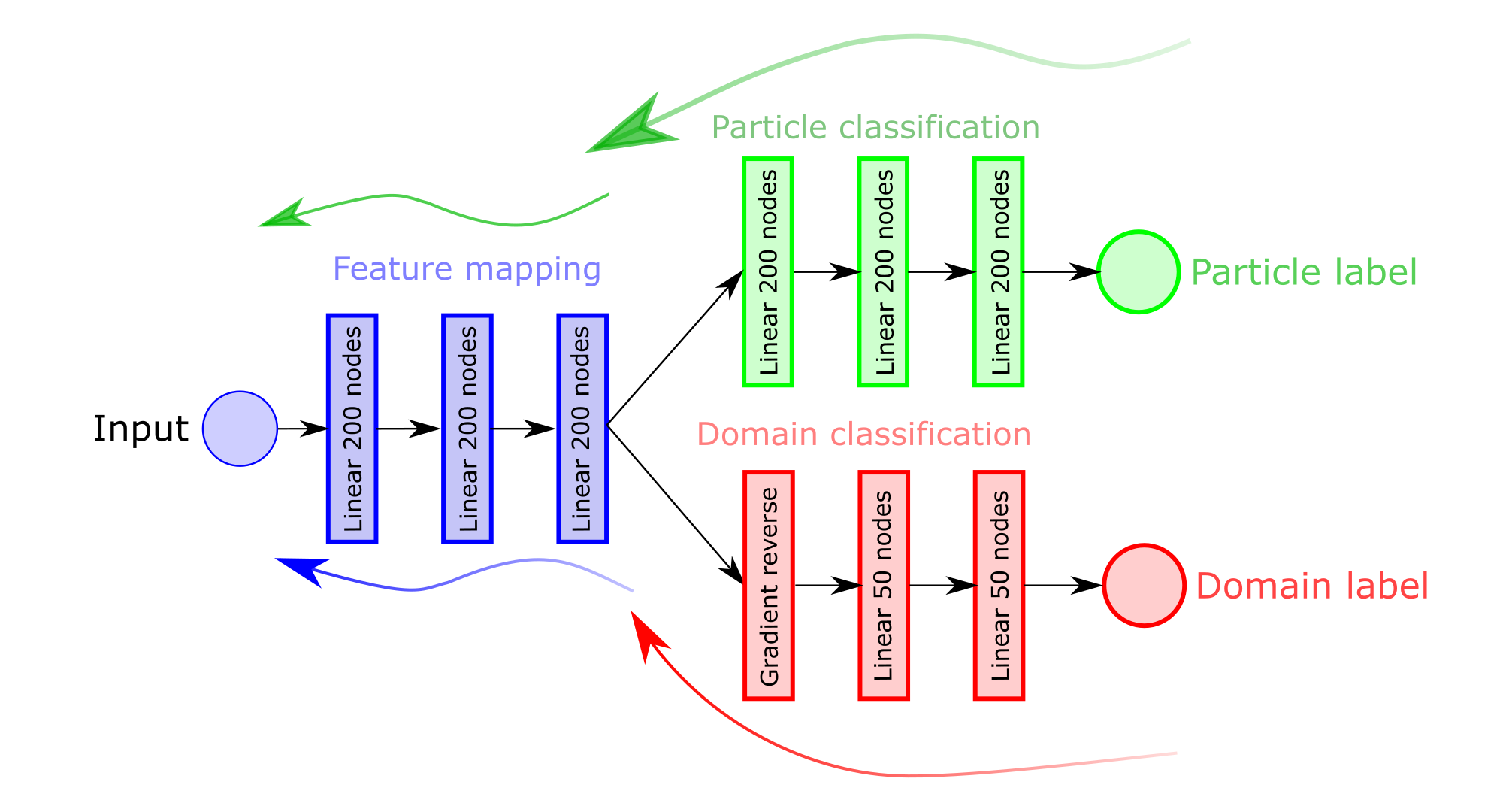}
    \caption{Architecture of Domain Adversarial Neural Network.}
    \label{fig:pid-dann}
\end{figure}

In the world of neural networks, Domain Adversarial Neural Network (DANN)~\cite{ganin2016domain} is the realization of the domain adaptation technique. As depicted in Figure~\ref{fig:pid-dann}, DANN is composed of three neural networks. The feature mapping module maps the original input into domain invariant features, which are provided to the particle classifier that outputs the particle type. At the same time, the domain classifier enforces domain invariance of extracted features through adversarial training.

Adversarial training requires DANN training to be split into two steps. First, the domain classifier takes current features from the feature mapping network and assigns them domain labels, whether the data comes from a real or a simulation source. Then, the weights of the domain classifier are frozen, and the particle classifier and the feature mapper learn jointly to predict accurate particle species. The weights of the particle classifier and the domain classifier are updated with respective gradients, while the feature mapper weights are updated with a gradient from the particle classifier and a reversed gradient from the domain classifier. As a result, the trained model maximizes particle identification scores while minimizing domain classifier scores to ensure that the domains are hardly distinguishable for the particle classifier. Overall, the training procedure is more complex than in a simple neural network, but the application performance is similar and depends on the complexity of the particle classifier and the feature mapper.

Domain adaptation is widely used in natural language processing~\cite{blitzer2007biographies,glorot2011domain} and computer vision~\cite{gopalan2011domain,fernando2013unsupervised}. In high-energy physics, the author of Ref.~\cite{walter2018domain} presents how this method can improve the quality of automatic jet tagging on real experimental data. The initial tests with DANN~\cite{JINST2022} show that this technique improves the classification of particles in experimental data.

\section{Conclusions and outlook}

Using machine learning for particle identification can result in higher purity and efficiency than standard methods. The current priority is to test PID ML in a real-world analysis of simulation data from the present Run 3 data-taking period. Afterward, we will extend the attention model with domain adaptation and test it on the new real data. We will check what levels of disagreement between experimental and Monte Carlo data DANN can handle, and we will estimate model uncertainties. Finally, we will be able to achieve regular production of models for Run 3.

In our project, we also solved a technical problem of integration of Python machine learning projects with the C++ ALICE software. Once a convenient data format conversion into ONNXRuntime tensors is developed, \osq ML projects will get an even more unified, user-friendly interface.

\acknowledgments

We would like to thank the ALICE Collaboration for guidance and support during our research as well as for the access to all software and data.

This work was supported by the Polish National Science Centre under agreements no. 2021/43/ D/ST2/02214 and UMO-2022/45/B/ST2/02029, by the Polish Ministry for Education and Science under agreements no. 2022/WK/01 and 5236/CERN/2022/0, as well as by the IDUB-POB-FWEiTE-2 project granted by Warsaw University of Technology under the program Excellence Initiative: Research University (ID-UB).


\bibliographystyle{JHEP}
\bibliography{biblio.bib}

\end{document}

%% file: commands.tex
\newcommand{\pt}           {\ensuremath{p_{\rm T}}\xspace}

\newcommand{\nineH}        {$\sqrt{s}=0.9$~Te\kern-.1emV\xspace}
\newcommand{\seven}        {$\sqrt{s}=7$~Te\kern-.1emV\xspace}
\newcommand{\twoH}         {$\sqrt{s}=0.2$~Te\kern-.1emV\xspace}
\newcommand{\twosevensix}  {$\sqrt{s}=2.76$~Te\kern-.1emV\xspace}
\newcommand{\five}         {$\sqrt{s}=5.02$~Te\kern-.1emV\xspace}
\newcommand{\thirteen}     {$\sqrt{s}=13$~Te\kern-.1emV\xspace}
\newcommand{\twosevensixnn}{$\sqrt{s_{\mathrm{NN}}}=2.76$~Te\kern-.1emV\xspace}
\newcommand{\fivenn}       {$\sqrt{s_{\mathrm{NN}}}=5.02$~Te\kern-.1emV\xspace}

\newcommand{\TeV}          {Te\kern-.1emV\xspace}
\newcommand{\GeV}          {Ge\kern-.1emV\xspace}
\newcommand{\MeV}          {Me\kern-.1emV\xspace}
\newcommand{\GeVmass}      {\ensuremath{\rm{Ge\kern-.1emV/}c^2\xspace}}
\newcommand{\MeVmass}      {\ensuremath{\rm{Me\kern-.1emV/}c^2\xspace}}
\newcommand{\GeVc}         {\ensuremath{\rm{Ge\kern-.1emV/}c\xspace}}
\newcommand{\MeVc}         {\ensuremath{\rm{Me\kern-.1emV/}c\xspace}}
\newcommand{\eVc}         {\ensuremath{\rm{e\kern-.1emV/}c\xspace}}
\newcommand{\GeVcc}        {\ensuremath{\rm{Ge\kern-.1emV/}c^2\xspace}}
\newcommand{\MeVcc}        {\ensuremath{\rm{Me\kern-.1emV/}c^2\xspace}}
\newcommand{\eVcc}        {\ensuremath{\rm{e\kern-.1emV/}c^2\xspace}}

\let\GeVmass=\GeVcc
\let\MeVmass=\MeVcc

\newcommand{\osq}         {\ensuremath{\rm{O}^2}\xspace}

%% file: main.bbl
\providecommand{\href}[2]{#2}\begingroup\raggedright\begin{thebibliography}{10}

\bibitem{Aamodt:2008zz}
{\scshape ALICE} collaboration, \emph{{The ALICE experiment at the CERN LHC}},
  \href{https://doi.org/10.1088/1748-0221/3/08/S08002}{\emph{JINST} {\bfseries
  3} (2008) S08002}.

\bibitem{Evans:2008zzb}
L.~Evans and P.~Bryant, \emph{{LHC Machine}},
  \href{https://doi.org/10.1088/1748-0221/3/08/S08001}{\emph{JINST} {\bfseries
  3} (2008) S08001}.

\bibitem{ALICE:2022wpn}
{\scshape ALICE} collaboration, \emph{{The ALICE experiment -- A journey
  through QCD}}, {\emph{arXiv:2211.04384 [nucl-ex]} (2022) }
  [\href{https://arxiv.org/abs/2211.04384}{{\ttfamily 2211.04384}}].

\bibitem{Tauro:2263642}
A.~Tauro, ``{ALICE Schematics}.'' 2017.

\bibitem{tpc-tdr}
{\scshape ALICE} collaboration, \emph{{ALICE Time Projection Chamber: Technical
  Design Report}}, Technical design report. ALICE, CERN, Geneva (2000).

\bibitem{tof-tdr}
{\scshape ALICE} collaboration, \emph{{ALICE Time-Of-Flight system (TOF):
  Technical Design Report}}, Technical design report. ALICE, CERN, Geneva
  (2000).

\bibitem{trd-tdr}
{\scshape ALICE} collaboration, \emph{{ALICE Transition-Radiation Detector:
  Technical Design Report}}, Technical design report. ALICE, CERN, Geneva
  (2001).

\bibitem{ALICE:2016zzl}
{\scshape ALICE} collaboration, \emph{{Particle identification in ALICE: a
  Bayesian approach}},
  \href{https://doi.org/10.1140/epjp/i2016-16168-5}{\emph{European Physics
  Journal Plus} {\bfseries 131} (2016) 168}
  [\href{https://arxiv.org/abs/1602.01392}{{\ttfamily 1602.01392}}].

\bibitem{lhcb2015lhcb}
{\scshape LHCb} collaboration, \emph{{{LHC}b detector performance}},
  \href{https://doi.org/10.1142/S0217751X15300227}{\emph{International Journal
  of Modern Physics A} {\bfseries 30} (2015) 1530022}.

\bibitem{collado2021learning}
J.~Collado et~al., \emph{{Learning to identify electrons}},
  \href{https://doi.org/10.1103/PhysRevD.103.116028}{\emph{Physical Review D}
  {\bfseries 103} (2021) 116028}.

\bibitem{cms2022identification}
{\scshape CMS} collaboration, \emph{{Identification of hadronic tau lepton
  decays using a deep neural network}},
  \href{https://doi.org/10.1088/1748-0221/17/07/P07023}{\emph{Journal of
  Instrumentation} {\bfseries 17} (2022) P07023}.

\bibitem{JINST2022}
{\L}.K.~Graczykowski, M.~Jakubowska, K.R.~Deja, M.~Karwowska and on~behalf
  of~the ALICE~collaboration, \emph{{Using machine learning for particle
  identification in ALICE}},
  \href{https://doi.org/10.1088/1748-0221/17/07/C07016}{\emph{Journal of
  Instrumentation} {\bfseries 17} (2022) C07016}.

\bibitem{gal2016dropout}
Y.~Gal and Z.~Ghahramani, \emph{{Dropout as a Bayesian approximation:
  Representing model uncertainty in deep learning}},  in \emph{International
  Conference on Machine Learning}, pp.~1050--1059, PMLR, 2016.

\bibitem{o2physics}
AliceO2Group, ``O2 analysis framework documentation.'' 2024,
  \url{https://aliceo2group.github.io/analysis-framework/} (Accessed: 15
  January 2024).

\bibitem{BRUN199781}
R.~Brun and F.~Rademakers, \emph{Root -- an object oriented data analysis
  framework},
  \href{https://doi.org/https://doi.org/10.1016/S0168-9002(97)00048-X}{\emph{Nuclear
  Instruments and Methods in Physics Research Section A: Accelerators,
  Spectrometers, Detectors and Associated Equipment} {\bfseries 389} (1997)
  81}.

\bibitem{onnx}
{ONNX Community}, ``{ONNX}.'' 2024, \url{https://onnx.ai/} (Accessed: 15
  January 2024).

\bibitem{tensorflow2015-whitepaper}
M.~Abadi et~al., \emph{{TensorFlow: Large-Scale Machine Learning on
  Heterogeneous Systems}},  2015.

\bibitem{pytorch}
A.~Paszke et~al., \emph{{PyTorch: An Imperative Style, High-Performance Deep
  Learning Library}},  in \emph{Advances in Neural Information Processing
  Systems}, vol.~32, pp.~8024--8035 (2019).

\bibitem{onnxruntime}
{ONNXRuntime Community}, ``{ONNXRuntime}.'' 2024, \url{https://onnxruntime.ai/}
  (Accessed: 15 January 2024).

\bibitem{Alkin:2021mfo}
A.~Alkin, G.~Eulisse, J.F.~Grosse-Oetringhaus, P.~Hristov and M.~Karwowska,
  \emph{{ALICE Run 3 Analysis Framework}},
  \href{https://doi.org/10.1051/epjconf/202125103063}{\emph{EPJ Web Conf.}
  {\bfseries 251} (2021) 03063}.

\bibitem{kasak2023machinelearningbased}
M.~Kasak, K.~Deja, M.~Karwowska, M.~Jakubowska, Łukasz Graczykowski and
  M.~Janik, \emph{{Machine-learning-based particle identification with missing
  data}},  2023.

\bibitem{wang2019attention}
Z.~Wang et~al., \emph{{Attention-based multi-instance neural network for
  medical diagnosis from incomplete and low quality data}},  in \emph{2019
  International Joint Conference on Neural Networks (IJCNN)}, pp.~1--8, 2019,
  \href{https://doi.org/10.1109/IJCNN.2019.8851846}{DOI}.

\bibitem{grangier2010feature}
D.~Grangier and I.~Melvin, \emph{{Feature set embedding for incomplete data}},
  {\emph{Advances in Neural Information Processing Systems} {\bfseries 23}
  (2010) }.

\bibitem{vaswani2017attention}
A.~Vaswani et~al., \emph{{Attention is all you need}}, {\emph{Advances in
  Neural Information Processing Systems} {\bfseries 30} (2017) }.

\bibitem{Sjostrand:2014zea}
T.~Sj\"ostrand et~al., \emph{{An introduction to PYTHIA 8.2}},
  \href{https://doi.org/10.1016/j.cpc.2015.01.024}{\emph{Computer Physics
  Communications} {\bfseries 191} (2015) 159}
  [\href{https://arxiv.org/abs/1410.3012}{{\ttfamily 1410.3012}}].

\bibitem{Brun:1994aa}
R.~Brun et~al., \emph{{GEANT Detector Description and Simulation Tool}},  Tech.
  Rep. CERN-W5013, CERN-W-5013, W5013, W-5013, CERN (1994),
  \href{https://doi.org/10.17181/CERN.MUHF.DMJ1}{DOI}.

\bibitem{ganin2016domain}
Y.~Ganin et~al., \emph{Domain-adversarial training of neural networks},
  \href{https://doi.org/10.1007/978-3-319-58347-1_10}{\emph{{The Journal of
  Machine Learning Research}} {\bfseries 17} (2016) 2096}.

\bibitem{blitzer2007biographies}
J.~Blitzer, M.~Dredze and F.~Pereira, \emph{{Biographies, Bollywood, Boom-boxes
  and Blenders: Domain Adaptation for Sentiment Classification}},  in
  \emph{Proceedings of the 45th Annual Meeting of the Association of
  Computational Linguistics}, pp.~440--447, 2007.

\bibitem{glorot2011domain}
X.~Glorot, A.~Bordes and Y.~Bengio, \emph{{Domain adaptation for large-scale
  sentiment classification: A deep learning approach}},  in \emph{ICML}, 2011.

\bibitem{gopalan2011domain}
R.~Gopalan, R.~Li and R.~Chellappa, \emph{{Domain adaptation for object
  recognition: An unsupervised approach}},  in \emph{2011 International
  Conference on Computer Vision}, pp.~999--1006, IEEE, 2011,
  \href{https://doi.org/10.1109/ICCV.2011.6126344}{DOI}.

\bibitem{fernando2013unsupervised}
B.~Fernando et~al., \emph{Unsupervised visual domain adaptation using subspace
  alignment},  in \emph{Proceedings of the IEEE International Conference on
  Computer Vision}, pp.~2960--2967, 2013,
  \href{https://doi.org/10.1109/ICCV.2013.368}{DOI}.

\bibitem{walter2018domain}
D.~Walter, \emph{{Domain Adaptation Studies in Deep Neural Networks for
  Heavy-Flavor Jet Identification Algorithms with the CMS Experiment}},
  {\emph{Master thesis} (2018) }.

\end{thebibliography}\endgroup
